\documentclass[12pt,preprint]{aastex}

\accepted{2003 August 22}
\begin{document}

\centerline {Erratum: "Oscillations in Arcturus from {\em WIRE} photometry" (ApJ, 591, L151 [2003])}

\vskip 0.4 cm

\author{Alon Retter, Timothy R. Bedding, Derek L. Buzasi, Hans Kjeldsen 
and L\'aszl\'o L. Kiss}

\vskip 0.4 cm

The value of the mode lifetime given in the first paragraph of the 
Discussion should be divided by 2$\pi$ and is therefore 2.0 days
rather than 13 days.

\end{document}